\documentclass[sigconf]{acmart}

\AtBeginDocument{%
  \providecommand\BibTeX{{%
    \normalfont B\kern-0.5em{\scshape i\kern-0.25em b}\kern-0.8em\TeX}}}

\usepackage{latexsym}
\usepackage{graphicx}
\graphicspath{{./images/}}
\usepackage{booktabs} 
\usepackage{color}  
\usepackage{amsmath}  
\usepackage{subcaption}
\usepackage{caption}
\usepackage{tikz}
\usepackage{colortbl} 
\usepackage{framed}
\usepackage{multirow}
\usepackage{multicol}
\usepackage{hyperref}
\usepackage{url}
\usepackage{balance}
\usepackage{verbatim}
\usepackage{cancel}
\usepackage{xspace} 
\usepackage[ruled,vlined]{algorithm2e}
\usepackage{bbold} 

\newcommand{\acc}{\textsc{Accent}\xspace}
\newcommand{\prince}{\textsc{Prince}\xspace}
\newcommand{\fia}{\textsc{FIA}\xspace}
\newcommand{\ncf}{\textsc{NCF}\xspace}
\newcommand{\rcf}{\textsc{RCF}\xspace}

\newcommand{\struct}[1]{\texttt{\small #1}}

\acmConference[SIGIR'21]{ACM SIGIR Conference}{July 2021}
{Montréal, Québec}
\acmYear{2021}
\copyrightyear{2021}

\acmArticle{4}
\acmPrice{15.00}

\begin{document}
\fancyhead{}

\title{Counterfactual Explanations for Neural Recommenders}

\author{Khanh Hiep Tran}
\affiliation{%
  \institution{MPI for Informatics, Germany}
}
\email{ktran@mpi-inf.mpg.de}

\author{Azin Ghazimatin}
\affiliation{%
  \institution{MPI for Informatics, Germany}
}
\email{aghazima@mpi-inf.mpg.de}

\author{Rishiraj Saha Roy}
\affiliation{%
  \institution{MPI for Informatics, Germany}
}
\email{rishiraj@mpi-inf.mpg.de}

\renewcommand{\shortauthors}{K. Tran et al.}

\newcommand{\squishlist}{
 \begin{list}{$\bullet$}
  { \setlength{\itemsep}{0pt}
     \setlength{\parsep}{1pt}
     \setlength{\topsep}{1pt}
     \setlength{\partopsep}{0pt}
     \setlength{\leftmargin}{1.5em}
     \setlength{\labelwidth}{1em}
     \setlength{\labelsep}{0.5em} } }

\newcommand{\squishend}{
  \end{list}  }

\begin{abstract}
Understanding why specific items are recommended to users can significantly increase their trust and satisfaction in the system. While neural recommenders have become the state-of-the-art in recent years, the complexity of deep models still makes the generation of tangible explanations for end users a challenging problem. Existing methods are usually based on attention distributions over a variety of features, which are still questionable regarding their suitability as explanations, and rather unwieldy to grasp for an end user. Counterfactual explanations based on a small set of the user's own actions have been shown to be an acceptable solution to the tangibility problem. However, current work on such counterfactuals cannot be readily applied to neural models. In this work, we propose ACCENT, the first general framework for finding counterfactual explanations for neural recommenders. It extends recently-proposed influence functions for identifying training points most relevant to a recommendation, from a single to a pair of items, while deducing a counterfactual set in an iterative process. We use ACCENT to generate counterfactual explanations for two popular neural models, Neural Collaborative Filtering (NCF) and Relational Collaborative Filtering (RCF), and demonstrate its feasibility on a sample of the popular MovieLens 100K dataset.
\end{abstract}

%
%




\maketitle

\section{Introduction}
\label{sec:introduction}

\begin{figure}[t]
	\centering
	\includegraphics[width=\columnwidth]{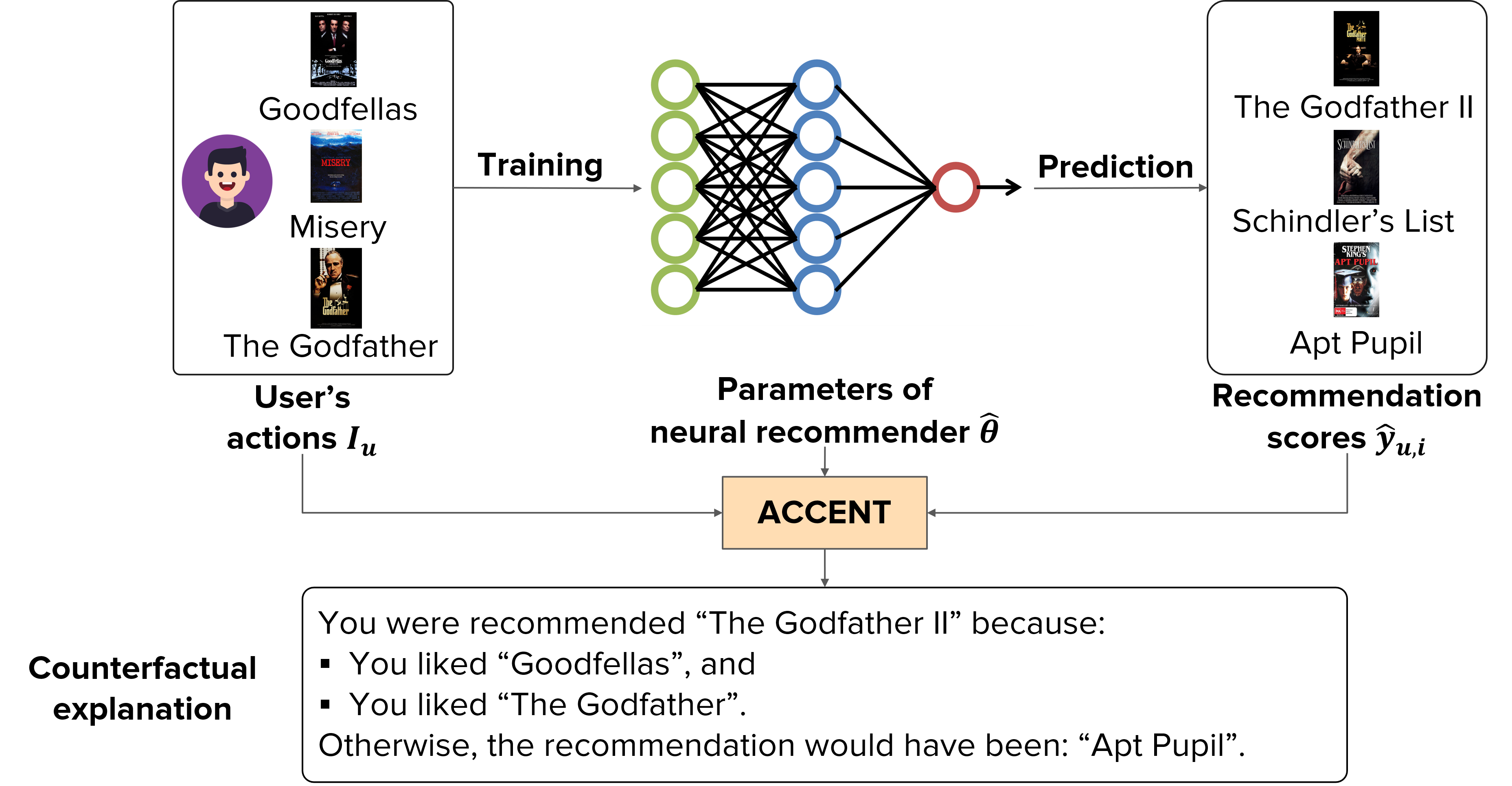}
	\caption{Positioning \acc in a neural recommender setup. Taking the user's actions, parameters of the neural recommender, and predicted scores for items as input, \acc produces a concise counterfactual explanation.}
	\label{fig:accent_ex}
	\vspace*{-0.3cm}
\end{figure}

\textbf{Motivation.}
Recommender systems have become ubiquitous in today's online world, spanning e-commerce to news to social media.
It is fairly well-accepted that high-quality explanations~\cite{ribeiro2016should,lundberg2017unified,chen2018learning} for the recommended content can help improve
users' satisfaction, while being actionable towards improving the underlying models~\cite{ghazimatin2021elixir,balog2020measuring,lu2018like,zhang2020explainable,zhao2019personalized,luo2020deep}.
Typical methods explaining neural recommenders face certain concerns:
(i) they often rely on the attention mechanism to find important words~\cite{seo2017interpretable}, reviews~\cite{chen2018neural}, or regions in images~\cite{chen2019personalized}, which is still controversial~\cite{jain2019attention,wiegreffe2019attention};
(ii) use connecting paths between users and items~\cite{yang2018towards,ai2018learning,xian2019reinforcement} that may not really be actionable and have privacy concerns; and,
(iii) they use 
external item metadata such as
reviews~\cite{seo2017interpretable,wu2019context,lu2018coevolutionary} or 
images~\cite{chen2019personalized,wu2019context}, that may not always be available.

In this context, it is reasonable to assume that in order to be tangible to end users, such explanations should relate to the user's own activity, and be scrutable, actionable, and concise~\cite{balog2019transparent,ghazimatin2019fairy}. This paved the way to posit \textit{counterfactual explanations} based on the user's own actions as a viable mechanism to address the tangibility concern~\cite{ghazimatin2020prince,karimi2021algorithmic,lucic2021cfgnnexplainer,mothilal2020explaining,martens2014explaining}. A counterfactual explanation is a set of the user's own actions, that, when removed, produces a different recommendation (referred to as a \textit{replacement item} in this text).
In tandem with the huge body of work on explanations, recommender models themselves have continued to become increasingly complex.
In recent years, neural 
recommender systems have become the de facto standard in the community,
owing to their power of learning the sophisticated non-linear interplay between several factors~\cite{zhang2019deep,he2017neural,xin2019relational}.
However, this same complexity prevents us from generating counterfactual explanations with the same methodology that works well for graph-based recommenders (the \prince algorithm~\cite{ghazimatin2020prince}).

\textbf{Approach.} To address this research gap, we present our method \acc (\underline{Ac}tion-based \underline{C}ounterfactual \underline{E}xplanations for \underline{N}eural Recommenders for \underline{T}angibility), that extends the basic idea in \prince to neural recommenders. However, this necessitates tackling two basic challenges: (i) \prince relied on estimating \textit{contribution scores} of a user's actions using Personalized PageRank for deriving counterfactual sets, something that does not carry over to arbitrary neural recommenders; and (ii) the graph-based \textit{theoretical formalisms} that form the core of the \prince algorithm, and ensure its optimality, also do not readily extend to deep learning models.
To overcome these obstacles, we adapt the recently proposed Fast Influence Analysis (FIA)~\cite{cheng2019incorporating} mechanism that sorts the user's actions based on their \textit{approximate influence} on the prediction from the neural recommender. 
While such influence scores are a viable proxy for the contribution scores above, they cannot be directly used to produce counterfactual sets. \acc extends the use of influence scores from single data points to \textit{pairs of items}, the pair being the recommendation item and its replacement. This is a key step that enables producing counterfactual sets 
by iteratively closing the score gap between the original recommendation and a candidate replacement item from the original top-$k$ recommendations.

Figure~\ref{fig:accent_ex} illustrates a typical counterfactual explanation output by \acc for the recommendation \struct{The Godfather II}:
had the user not watched movies \struct{Goodfellas} and \struct{The Godfather}, 
she would have been recommended \struct{Apt Pupil} instead.
More formally, given a user $u$, a list $I_u$ of items interacted by $u$ (her own actions), and a recommendation $rec$, 
\acc finds a counterfactual explanation $I_u^* \subseteq I_u$ whose removal from the training set results in a different recommendation $rec^*$.
We apply \acc to explain predictions from two prominent neural recommenders, Neural Collaborative Filtering (NCF)~\cite{he2017neural} and Relational Collaborative Filtering (RCF)~\cite{xin2019relational}, and validate it on a subset of the MovieLens 100K dataset. 
All code and data are available at \url{https://www.mpi-inf.mpg.de/impact/counterfactual-explanations-for-recommenders} and \url{https://github.com/hieptk/accent}.





\section{Method}
\label{sec:method}

\subsection{Estimating parameters}
\label{subsec:params}

Rooted in statistics, 
influence functions 
estimate how the model parameters change when a data point is upweighted by a small amount $\epsilon$~\cite{hampel1974influence}.
Using influence functions, Koh and Liang~\cite{koh2017understanding} proposed a 
method for estimating the impact of removing a data point 
from the training set (reducing its weight to $0$) on the model parameters. 
To briefly describe this method, 
let $\hat{\theta}$ be 
the original model parameters, i.e., $\hat{\theta}$ is the minimizer of the empirical risk:
\begin{align}
R(\theta) = \frac{1}{n} \sum_{i=1}^n L(z_i, \theta)
\end{align}
where $n$ is the number of training points and $L(z_i, \theta)$ is the loss of training point $z_i$ using parameters $\theta$. After upweighting training point $z$ by a small amount $\epsilon$, the new optimal parameters ($\hat{\theta}_{\epsilon, z}$) are 
approximated as follows:
\begin{align}
\hat{\theta}_{\epsilon, z} = \arg \min_{\theta \in \Theta} \{R(\theta) + \epsilon L(z, \theta)\}
\end{align}
According to~\cite{koh2017understanding}, the influence of upweighting training point $z$ by a small amount $\epsilon$ is given by:
\begin{align}
\label{eq:param_deriv}
\frac{d\hat{\theta}_{\epsilon, z}}{d\epsilon} \Big|_{\epsilon=0} = -H_{\hat{\theta}}^{-1} \nabla L(z, \hat{\theta})
\end{align}
where $H_{\hat{\theta}}$ is the Hessian matrix
computed as $H_{\hat{\theta}} = \frac{1}{n} \sum_{i=1}^n \nabla_\theta^2 L(z_i, \hat{\theta})$.
Since the weight of each training point is $\frac{1}{n}$, removing a training point $z$ is equivalent to upweighting it by $-\frac{1}{n}$. We can estimate the new parameters $\hat{\theta}^{-z}$ after removing $z$ by setting $\epsilon = -\frac{1}{n}$ in (\ref{eq:param_deriv}):
\begin{align}
\hat{\theta}^{-z} - \hat{\theta} = \frac{1}{n} H_{\hat{\theta}}^{-1} \nabla L(z, \hat{\theta})
\end{align}
Note that, since neural models are not convex, we enforce the invertibility of $H_{\hat{\theta}}$ by adding a small damping term $\lambda$ to its diagonal. \par
A challenge while applying influence functions to neural models is that the set of parameters is usually huge. FIA~\cite{cheng2019incorporating} is a technique to reduce the computational cost of influence functions for recommender systems, and we will adapt this approach to our algorithm.

\subsection{Computing influence on score gaps}
\label{subsec:gaps}

Using FIA, we can estimate the new parameters $\hat{\theta}^{-z}$ when one training point (one of the user's actions $I_u$) is removed. 
Substituting $\hat{\theta}^{-z}$ into the recommender model, we can estimate the new predicted score for any item.
We denote the preference score of user $u$ for item $i$ before and after removing point $z$ 
as $\hat{y}_{u,i}$ and $\hat{y}_{u,i}^{-z}$, respectively. The influence of $z$ on $\hat{y}_{u,i}$ is defined as:
\begin{align}
I(z, \hat{y}_{u,i}) = \hat{y}_{u,i} - \hat{y}_{u,i}^{-z}
\end{align}
We can then estimate the influence of a training point $z$ on the \textit{score gap} between two items $i$ and $j$ as:
\begin{align}
&I(z, \hat{y}_{u,i} - \hat{y}_{u,j}) = (\hat{y}_{u,i} - \hat{y}_{u,j}) - (\hat{y}_{u,i}^{-z} - \hat{y}_{u,j}^{-z}) \nonumber \\
&= (\hat{y}_{u,i} - \hat{y}_{u,i}^{-z}) - (\hat{y}_{u,j} - \hat{y}_{u,j}^{-z}) = I(z, \hat{y}_{u,i}) - I(z, \hat{y}_{u,j})
\end{align} 
For a set of points $Z = \{z_1, z_2, ..., z_m\}$, we approximate the influence of removing this set by:
\begin{align}
I(Z, \hat{y}_{u,i} - \hat{y}_{u,j}) = \sum_{k=1}^m I(z_k, \hat{y}_{u,i} - \hat{y}_{u,j})
\end{align}
With the increase in the size of the set $Z$, the accuracy of the above estimation deteriorates. 
However, since counterfactual sets are usually very small, this approximation is still valid.

\subsection{Filling the gap}
\label{subsec:fill}

To replace the recommendation $rec$ with $rec^*$, we need to find a counterfactual explanation set 
$Z \subseteq I_u$ whose removal results in: 
\begin{align}
& \hat{y}^{-Z}_{u,rec} - \hat{y}^{-Z}_{u,rec^*} < 0 \nonumber \\
& \Leftrightarrow \hat{y}_{u,rec} - \hat{y}_{u,rec^*} - \hat{y}^{-Z}_{u,rec} + \hat{y}^{-Z}_{u,rec^*} > \hat{y}_{u,rec} - \hat{y}_{u,rec^*} \nonumber \\
& \Leftrightarrow I(Z, \hat{y}_{u,rec} - \hat{y}_{u,rec^*}) > \hat{y}_{u,rec} - \hat{y}_{u,rec^*} \nonumber \\
& \Leftrightarrow \sum_{k=1}^m I(z_k, \hat{y}_{u,rec} - \hat{y}_{u,rec^*}) > \hat{y}_{u,rec} - \hat{y}_{u,rec^*} \label{eq:swap_cond}
\end{align}
Therefore, the optimal way to replace $rec$ with $rec^*$ is to add training points $z_k$ to $Z$ in the order of decreasing $I(z_k, \hat{y}_{u,rec} - \hat{y}_{u,rec^*})$ until (\ref{eq:swap_cond}) is satisfied. To find the smallest counterfactual explanation $I_u^*$, we try every replacement item from a set of candidates $I_{rep}$. In principle, $I_{rep}$ could span the complete set of items $I$, but a practical choice is the original set of top-$k$ recommendations. Good models usually diversify their top-$k$ while preserving relevance, and choosing the replacement from this slate ensures that $rec^*$ is neither trivially similar to $rec$ nor irrelevant to $u$. Finally, this smallest set of actions $I_u^*$ is returned to $u$ as a tangible explanation for 
$rec$. \textbf{Algorithm \ref{algo:accent}} contains a precise formulation of \acc. \acc's time complexity is $O(|I_{rep}| \times |I_u| \times \log |I_u|)$ + $O(|I_{rep}| \times |I_u|)$ calls of FIA. Since \acc only requires access to gradients and the Hessian matrix, it is applicable to a large class of neural recommenders.

\setlength{\textfloatsep}{0pt} 
\begin{algorithm}[t]
    \small
	\textbf{Input:} user $u$, recommendation item $rec$, \\
	items interacted with by user $I_u$, candidate replacement items $I_{rep}$ \\
	\textbf{Output:} smallest counterfactual set $I_u^*$, replacement $rec^*$ \\
	\SetAlgoLined
	\DontPrintSemicolon
	$I_u^*, rec^* \leftarrow I_u, -1$\;
	\For{$i \in I_{rep}$} {
		\For(\tcp*[f]{Compute influence on gap}){$z \in I_u$} {
			\textbf{compute} $I(z, \hat{y}_{u,rec})$ and $I(z, \hat{y}_{u,i})$\;
			$I(z, \hat{y}_{u,rec} - \hat{y}_{u,i}) \leftarrow I(z, \hat{y}_{u,rec}) - I(z, \hat{y}_{u,i})$\;
		}
		$gap, I_u^i \leftarrow \hat{y}_{u,rec} - \hat{y}_{u,i}, \emptyset$\;
		\textbf{sort} $I_u$ by decreasing $I(z, \hat{y}_{u,rec} - \hat{y}_{u,i})$\;
		\For{$z \in sorted(I_u)$} {
			\If{$gap < 0 $ \textbf{or} $ I(z, \hat{y}_{u,rec} - \hat{y}_{u,i}) \leq 0$} {
				\tcp*[l]{Gap is filled or impossible}
				\textbf{break}\;
			}
			$gap \leftarrow gap - I(z, \hat{y}_{u,rec} - \hat{y}_{u,i})$ \tcp*{Update gap}
			$I_u^i \leftarrow I_u^i \cup \{z\}$ \tcp*{and result set}
		}
		\If(\tcp*[f]{New smallest set}){$gap < 0$ \textbf{and} $|I_u^i| < |I_u^*|$} {
			$I_u^*, rec^* \leftarrow I_u^i, i$\;
		}
	}
	\Return $I_u^*, rec^*$
	\caption{\acc}
	\label{algo:accent}
\end{algorithm}

\section{Experimental Setup}
\label{sec:setup}

\subsection{Recommender models}
\label{subsec:rec-models}

We apply \acc on 
\ncf~\cite{he2017neural}, one of the first neural recommenders,  
and \rcf~\cite{xin2019relational}, a more recent choice.  
\ncf~\cite{he2017neural} consists of
a generalized matrix factorization layer (where the user and the item embeddings are element-wise multiplied), and a multilayer perceptron that takes these user and item embeddings as input. These two parts are then combined to predict the final recommendation score. 
\rcf 
uses auxiliary information to incorporate item-item relations into the model.
It computes target-aware embeddings
that capture information about the user, her interactions, and their relationships with target items (recommendation candidates) using a two-layer attention scheme. The recommendation score of the user and the target item are computed from these target-aware embeddings.

\subsection{Dataset}
\label{subsec:data}

We
use the popular MovieLens $100$K dataset~\cite{harper2015movielens}, 
which contains $100k$ ratings on a $1-5$ scale 
by $943$ users on $1682$ movies. 
Input of this form can be directly fed into \ncf. On the other hand, to conform to the implicit feedback setting in \rcf, we binarized ratings to a positive label if it is $3$ or above, and a negative label otherwise. We removed all users with $<\!10$ positive ratings or $<\!10$ negative ratings so that the profiles are big and balanced enough for learning discriminative user models. 
This pruning results in $452$ users, $1654$ movies, and $61054$ interactions in
our dataset. For item-item relations in \rcf, we used the auxiliary data provided in~\cite{xin2019relational}, which contains four relation types and $97209$ relation pairs.

\subsection{Baselines}
\label{subsec:baselines}

We compare \acc against 
four baseline algorithms. Two baselines are based on the attention weights in \rcf and are not applicable to \ncf, while the other two algorithms are based on \fia scores and can be used for both recommender models.

\subsubsection{Attention-based algorithms}
\label{attentionAlgo}
Attention weights in \rcf can be used to produce explanations~\cite{xin2019relational}. 
An item's attention weight shows how much it affects the prediction: thus, to find a counterfactual explanation for $rec$, 
we can sort all items in $I_u$ by decreasing attention weights. We then add these items one by one to $I_u^*$ until the predicted recommendation is changed.
Here, we assume removing a few training points does not change the model significantly, so all parameters remain fixed. We refer to this as \textbf{pure attention}.
%
We adapt pure attention to a smarter \textbf{attention} baseline,
where an item is added to $I_u^*$ only if removing it reduces the score gap between $\hat{y}_{u,rec}$ and the second-ranked item's score. The underlying intuition is to avoid adding potentially irrelevant items to the explanation. The score gap is again estimated using fixed parameters.

\subsubsection{\fia-based algorithms}
Here, we test the direct applicability of \fia to produce counterfactual explanations.
We simply sort items in $I_u$ by $I(z, \hat{y}_{u,rec})$ and add items one by one to $I_u^*$ until $rec$ is displaced (\textbf{pure \fia}).
As with attention, we improve pure \fia to keep only the interactions that reduce the score gap between $\hat{y}_{u,rec}$ and the score of the second-ranked item (strategy denoted by \textbf{\fia}). 

\subsection{Initialization}
\label{subsec:init}

For \fia on \ncf, we used the implementation in~\cite{cheng2019incorporating}. We used a batch size $1246$ as this implementation requires 
this value to 
be a factor of 
the dataset size ($61054$). All other hyperparameters were kept the same. For \rcf, we set the dropout rate to $0$ to minimize randomness during retraining. We replaced the ReLU activation with GELU~\cite{hendrycks2020gaussian} to avoid problems with non-differentiability~\cite{koh2017understanding}. To guarantee \fia's effectiveness, we made sure that each interaction corresponds to one training point (that was fifty in the original model). 
For this, we paired  
each 
liked item $i^+$ by user $u$ 
with one of her disliked items $i^-$, and   
added triples $(u,i^+,i^-)$ to the training set. 
In particular, for each $i^+$, we selected an $i^-$ that shares the highest number of relations with $i^+$. By doing this principled negative sampling, the RCF model can still discriminate between positive and  negative items effectively, despite having only one negative item for each positive. For \fia on \rcf, we added damping term $\lambda = 0.01$ to the Hessian matrix and 
used our own implementation. 




\section{Results and insights}
\label{sec:results}

\subsection{Evaluation protocol}
\label{subsec:protocol}

For each of the $452$ users in our dataset, we find an explanation $I_u^*$ 
for their recommendation $rec$, and a replacement $rec^*$ from $I_{rep}$, where
$I_{rep}$ is the original top-$k$ ($k = 5, 10, 20$). We then retrain the models without $I_u^*$ and verify if $rec^*$ replaces $rec$. 
This is done for both recommender models (\ncf and \rcf) and each of the explanation algorithms (\acc, pure attention, attention, pure \fia and \fia, as applicable).
The percentages of 
actual replacements  (CF percentage)
and the average sizes of the 
counterfactual sets (CF set size)
for \acc and the baselines
are reported in Table~\ref{tab:main-res}. Ideally, an algorithm should have a high CF percentage and a small CF set size. To give a qualitative feel of the explanations generated by \acc, we provide some anecdotal examples in Table~\ref{tab:anecdotes} (baselines had larger CF sets). 
To compare the counterfactual effect (CF percentages) between two methods, we used the McNemar's test for paired binomial data, since each explanation 
is either actually counterfactual or not (binary).  
For CF set sizes, we used the one-sided paired $t$-test. The significance level for all tests was set to $0.05$.



\begin{table*} [t] \small
	\newcolumntype{G}{>{\columncolor [gray] {0.90}}c}
	\begin{tabular}{l l c c c c c c}
		\toprule
		\multicolumn{2}{l}{\textbf{Candidate top-\textit{k} set of replacement items}} & \multicolumn{2}{c}{\textbf{$k=5$}} & \multicolumn{2}{c}{\textbf{$k=10$}} & \multicolumn{2}{c}{\textbf{$k=20$}} \\		
		\cmidrule(lr){1-2}
		\cmidrule(lr){3-4}
		\cmidrule(lr){5-6}
		\cmidrule(lr){7-8}
		 \textbf{Recommender model} & \textbf{Explanation model} & \textbf{CF percentage} & \textbf{CF set size} & \textbf{CF percentage} & \textbf{CF set size} & \textbf{CF percentage} & \textbf{CF set size} \\
		\toprule
		\textbf{NCF} \cite{he2017neural} & \textbf{Pure FIA} \cite{cheng2019incorporating} & $54.20$ & $9.08$ & $56.19$ & $9.46$ & $55.75$ & $9.50$ \\
		& \textbf{FIA} \cite{cheng2019incorporating} & $55.97$ & $7.98$ & $56.19$ & $7.80$ &  $55.75$ & $7.84$ \\
		& \textbf{ACCENT (Proposed)} & $\boldsymbol{57.30}$ & $\boldsymbol{4.73^*}$ & $\boldsymbol{57.74}$ & $\boldsymbol{4.69^*}$ & $\boldsymbol{57.08}$ & $\boldsymbol{4.62^*}$ \\
		\midrule
		\textbf{RCF} \cite{xin2019relational} & \textbf{Pure Attention} \cite{xin2019relational}& $73.01$ & $9.36$ & $73.45$ & $7.94$ & $74.34$ & $7.75$ \\
		& \textbf{Attention} \cite{xin2019relational} & $76.99$ &	$3.55$ & $76.99$ & $3.53$ & $76.99$ & $3.51$ \\
		& \textbf{Pure FIA} \cite{cheng2019incorporating} & $80.75$ & $4.85$ & $81.19$ & $4.62$ & $81.86$ & $4.72$ \\
		& \textbf{FIA} \cite{cheng2019incorporating} & $81.64$ & $4.15$ & $81.86$ & $4.10$ & $81.86$ & $4.10$ \\
		& \textbf{ACCENT (Proposed)} & $\boldsymbol{81.86^\dagger}$ & $\boldsymbol{2.83^{*\dagger}}$ & $\boldsymbol{82.08^\dagger}$ & $\boldsymbol{2.75^{*\dagger}}$ & $\boldsymbol{82.08^\dagger}$ & $\boldsymbol{2.74^{*\dagger}}$ \\    \bottomrule
	\end{tabular}
	\\ \raggedright \small Best values in each column are in \textbf{bold}. * and $^\dagger$ denote statistical significance of \acc over FIA and Attention, respectively.
	\caption{Performance comparison of \acc with baselines on our sample of the MovieLens 100K benchmark.}
	\label{tab:main-res}
	\vspace*{-0.7cm}
\end{table*}

\begin{table}[t] \small
	\resizebox{\linewidth}{!} {
		\begin{tabular}{l l l}
			\toprule
			\textbf{Recommendation} & \begin{tabular}{l}\textbf{\acc Explanation}\end{tabular} & \textbf{Replacement} \\
			\midrule
			\struct{The Silence Of The Lambs} & \begin{tabular}{l}\struct{Contact}\\\struct{Fargo}\end{tabular} & \struct{Donnie Brasco} \\ \midrule
			\struct{Titanic} & \begin{tabular}{l}\struct{True Romance}\\ \struct{The Basketball Diaries} \end{tabular} & \struct{East Of Eden} \\ \midrule
			\struct{The Devil's Advocate} & \begin{tabular}{l}\struct{Speed}\\\struct{Eraser}\\\struct{It's A Wonderful Life}\end{tabular} & \struct{My Fair Lady} \\ 
			\bottomrule 
		\end{tabular} }
	\caption{Counterfactual sets generated by \acc.}
	\label{tab:anecdotes}
	\vspace{-0.3cm}
\end{table}

\subsection{Key findings}
\label{subsec:key}

\noindent \textbf{\acc is effective.} For both models, \acc produced the best overall results for producing counterfactual explanations.
Results are statistically significant for CF percentages over attention baselines, and for CF set sizes over both attention and FIA methods (marked with asterisks and daggers in Table~\ref{tab:main-res}).

\noindent \textbf{Attention is not explanation.} The two algorithms using attention performed the worst. Their CF percentage 
is at least $5\%$ lower than \acc
and
their average CF set sizes are between $1.3$ to $3.3$ times bigger than \acc.
This shows that using attention is not really helpful in finding concise counterfactual explanations. 

\noindent \textbf{FIA
is not enough.} The two algorithms that directly use \fia to rank interactions produced very big explanations. The average size of \fia's explanations is $1.5$ to $1.7$ times bigger than that of \acc (about twice as big for pure FIA with NCF, the context in which FIA was originally proposed).
This provides evidence that to replace $rec$, the influence on the score gap is more important than the influence on the score of $rec$ alone. 

\noindent \textbf{Considering the score gap is essential.} The two pure algorithms that do not consider the score gap between $rec$ and the replacement while expanding the explanation, performed worse than their smarter versions that do take this gap reduction into account. The difference can be as large as $4\%$ in CF percentages, 
with up to $2.6$ times bigger CF sets 
(pure attention, $k=5$).

\noindent \textbf{Influence estimation for sets is adequate.} Our approximation of \fia for sets of items is actually quite close to the true influence. 
For \rcf, the RMSE between the approximated influence and the true influence is $1.17$ over different values of $k$, which is small compared to the standard deviation of the true influence ($\simeq 3.7$). For \ncf, this RMSE is $0.36$ while the true influence has a standard deviation of $0.34$, implying that estimation accuracy is lower for this model: this in turn results in a lower CF percentage.

\noindent \textbf{Explanations get smaller as $k$ grows.} The performance of \acc is stable across different $k \; (5, 10, 20)$, varying less than $1\%$. 
The average CF set size slightly decreases as $k$ increases, because we have more options to replace $rec$ with. Interestingly, a similar effect was observed in the graph-based setup in \prince~\cite{ghazimatin2020prince}.

\vspace*{-0.2cm}
\subsection{Analysis}
\label{subsec:analysis}

\textbf{Pairwise vis-\`{a}-vis one-versus-all.} In our main algorithm, instead of fixing one replacement item at a time (pairwise), we can have a different approach that does not need a fixed replacement. In particular, at each step, we can reduce the gap between $rec$ and the second-ranked item at the time, leaving this second-ranked item to change freely during the process (one-versus-all). Through experiments, we found that this approach can slightly improve the counterfactual percentage of \acc by $0.22\%$ on RCF but at the cost of bigger explanations.

\noindent \textbf{Error analysis.} Despite being effective in estimating influence, \fia is still an approximation.
In particular, 
it assumes that only a few parameters are affected by the removal of a data point (corresponding user and item embeddings). 
This assumption can sometimes lead to errors in practice. 
In NCF, 
we observed a large discrepancy between the estimated influence and the actual score drops, 
despite their strong correlation ($\rho=0.77$). 
This explains \acc's relatively low CF percentages in \ncf ($\simeq 57\%$). 
It would thus be desirable to update more parameters when one action is removed:
for example, in RCF, we can also consider relation type and relation value embeddings. However, this could substantially
increase the computational cost. 
Another source of error is that the influence of a set of items is sometimes overestimated. This can cause \acc to stop prematurely when the cumulative influence is not enough to swap two items.
To mitigate this,
we can retrain the model after \acc stops, to verify whether the result is actually counterfactual. If not, \acc can resume and add more actions. 

\vspace*{-0.2cm}
\section{Conclusion}
\label{sec:conclusion}

We described \acc, a mechanism for generating counterfactual explanations for neural recommenders. \acc extends ideas from the \prince algorithm to the neural setup, while using and adapting influence values from FIA for the pairwise contribution scores that were a core component of \prince but non-trivial to obtain in deep models. We demonstrated \acc's effectiveness over attention and \fia baselines with the underlying recommender being NCF or RCF, but it is applicable to a much broader class of models: the only requirements are access to gradients and the Hessian. 

\noindent \textbf{Acknowledgements.}	This work was supported by the ERC Synergy Grant 610150 (imPACT).

\clearpage

\bibliographystyle{ACM-Reference-Format}
\balance
\bibliography{accent}

\end{document}